\documentclass[%
reprint,
superscriptaddress,
 amsmath,amssymb,
 aps,
pra,
]{revtex4-2}
\usepackage{graphicx}
\usepackage{dcolumn}
\usepackage{bm} 
\usepackage{braket}
\usepackage{tikz}
\usepackage{quantikz}
\usepackage{subcaption}
\usepackage[hidelinks]{hyperref}

\definecolor{red}{RGB}{0,0,0}
\definecolor{blue}{RGB}{0,0,0}
\definecolor{orange}{RGB}{0,0,0}
\definecolor{purple}{RGB}{152,78,163}

\let\vec\boldsymbol

\begin{document}

\title{Quantum Algorithm for the Advection-Diffusion Equation by \\ Direct Block Encoding of the Time-Marching Operator} 

\author{Paul Over}
\affiliation{%
Institute for Fluid Dynamics and Ship Theory, Hamburg University of Technology, Hamburg D-21073, Germany.
}%
\author{Sergio Bengoechea}
\affiliation{%
Institute for Fluid Dynamics and Ship Theory, Hamburg University of Technology, Hamburg D-21073, Germany.
}%
\author{Peter Brearley}
\affiliation{%
Department of Aeronautics, Imperial College London, London SW7 2AZ, UK.
}%
\author{Sylvain Laizet}%
\affiliation{%
Department of Aeronautics, Imperial College London, London SW7 2AZ, UK.
}%
\author{Thomas Rung}
\affiliation{%
Institute for Fluid Dynamics and Ship Theory, Hamburg University of Technology, Hamburg D-21073, Germany.
}%

\date{April 28, 2025}

\begin{abstract}
    A quantum algorithm for simulating multidimensional scalar transport problems using a time-marching strategy is presented. A direct unitary block encoding of the explicit time-marching operator is constructed, resulting in the intrinsic success probability of the squared solution norm without the need for amplitude amplification, thereby retaining a linear dependence on the simulation time. The algorithm separates the explicit time-marching operator into an advection-like component and a corrective shift operator. The advection-like component is mapped to a Hamiltonian simulation and combined with the shift operator through the linear combination of unitaries algorithm. State-vector simulations of a scalar transported in a steady two-dimensional Taylor-Green vortex support the theoretical findings.
\end{abstract}

\maketitle

\section{Introduction}

Partial differential equations (PDEs) provide a versatile mathematical framework for describing and simulating diverse phenomena, supporting technological progress with reduced costs compared to experimental methods. The computational demand for solving PDEs requires the extensive use of high-performance computing for many practical applications, leading to substantial investments in both hardware and algorithmic development. Despite these advancements, computational resources remain insufficient to address the scales of problems that scientists and engineers wish to simulate, often by several orders of magnitude \cite{Nasa2030}. Fault-tolerant quantum computing brings a paradigm shift in algorithmic performance by offering an exponential vector space for computation, although requiring the development of specialized algorithms to achieve this potential.

A prominent PDE in fluid dynamics is the advection-diffusion equation, given by
\begin{equation}
    \frac{\partial \phi}{\partial t} + \vec{v} \cdot \nabla \phi = D\nabla^2 \phi \, ,
    \label{eq:advection_diffusion_equation}
\end{equation}
that describes the transport of a scalar $\phi$ advected by an incompressible flow with velocity $\vec{v}$ in a medium with diffusivity $D$. The prevailing strategy for solving such PDEs on quantum computers is by conversion to a system of ordinary differential equations
\begin{equation}
    \frac{d\vec{\phi}}{dt} = M\vec{\phi}\, ,
    \label{eq:ODEs}
\end{equation}
applying the finite difference method for spatial discretization \cite{Berry2017, Krovi2023, Berry2024, An2023, Novikau2024, Jin2023, Hu2024}. This has the analytical solution of $\vec{\phi}(t) = e^{Mt}\vec{\phi}(0)$ for a matrix $M$ describing the discretized physics. To solve this more general problem, \citeauthor{Berry2017}~\cite{Berry2017} encoded a truncated Taylor series expansion of $e^{Mt}$ into a system of linear equations, and applied a quantum linear systems algorithm \cite{Harrow2009, Costa2022}. This has steadily been improved in subsequent works with the extension to a wider range of matrices \cite{Krovi2023} and with improved dependence on the precision \cite{Berry2024}. Other innovative approaches include the linear combination of Hamiltonian simulation method~\cite{An2023, Novikau2024} and Schr\"odingerization~\cite{Jin2023, Hu2024}, which map Eq.~\eqref{eq:ODEs} to a dilated system of Schr\"odinger equations that can be solved using Hamiltonian simulation. 

A less common strategy for solving problems in the form of Eq.~\eqref{eq:ODEs} on a quantum computer is time marching, where $e^{Mt}$ is approximated by its truncated Taylor series and evolved over a series of short time steps $\Delta t$ where the approximation is accurate. For example, the forward Euler method is accurate to first-order terms and results in the system
\begin{equation}
    \vec{\phi}(t+\Delta t) = A\vec{\phi}(t) \,,
    \label{eq:explicit_time_stepping}
\end{equation}
where $A = I + M\Delta t$ is the Euler time marching matrix and $I$ is the identity matrix. Time marching is often the most natural strategy for solving PDEs classically, so its limited use in quantum algorithms may be surprising. Since $A$ is non-unitary, it cannot be implemented directly as a series of quantum gates, instead requiring a probabilistic implementation by block encoding. Considering the following unitary block encoding implementation of Eq.~\eqref{eq:explicit_time_stepping},
\begin{equation}
    \begin{bmatrix}
        \frac{A}{\alpha} & * \\
        * & *
    \end{bmatrix}
    \begin{bmatrix}
        \ket{\phi_t} \\
        0
    \end{bmatrix} =
    \begin{bmatrix}
        \frac{A}{\alpha}\ket{\phi_t} \\
        *
    \end{bmatrix}\,,
    \label{eq:block_encoding}
\end{equation}
$A$ can only be encoded accurately up to a subnormalization factor $\alpha \ge 1$, where a subscript $t$ refers to a time index and the asterisk indicates arbitrary,  inconsequential blocks. Measuring $\ket{0}$ prepares the state $A/\alpha \ket{\phi_t}$ with probability $1/\alpha^2 \| A\ket{\phi_t} \|^2$. The constant factor of $1/\alpha^2$ induces an exponentially diminishing success probability with repeated time steps for $\alpha > 1$~\cite{Mezzacapo2015, Bharadwaj2024, Sanavio2024}, limiting the usefulness of quantum time-marching algorithms to $\alpha=1$.

Several algorithms capable of producing a general block encoding have been proposed \cite{Childs2012, Low2019, Gilyen2019}. For example, the linear combination of unitaries (LCU) algorithm \cite{Childs2012} is based on the principle that any non-unitary operator can be written as a weighted sum of unitary operators $A = \sum_j \beta_jU_j$. However, finding an efficient decomposition is not straightforward. Any non-unitary matrix can be written as the limit of the linear combination of four unitary operators
\begin{equation}
    A = \lim_{\tau\to 0} \frac{1}{2\tau} \left( ie^{-i\tau H_1} - ie^{i\tau H_1} +  e^{i\tau H_2} - e^{-i\tau H_2} \right) \, ,
    \label{eq:general}
\end{equation}
where $H_1 = (A+A^\dagger)/2$ and $H_2 = (A-A^\dagger)/2i$ are Hermitian matrices such that $A = H_1 + iH_2$. This methodology was used by \citeauthor{Bharadwaj2024}~\cite{Bharadwaj2024} in the LCU framework, but has the disadvantage that as $\tau\to 0$, $\alpha \to \infty$ in Eq.~\eqref{eq:block_encoding}, requiring a compromise between the accuracy and probability of success. Since the methodology cannot produce an accurate block encoding where $\alpha=1$, time marching results in an exponentially decaying probability of success with the simulation time $T$. This problem can be overcome by the quantum singular value transformation \cite{Gilyen2019}, which unifies qubitization \cite{Low2019} and quantum signal processing \cite{Low2017} algorithms for performing arbitrary polynomial transformations on the singular values of a block-encoded matrix. This was the approach taken by \citeauthor{Fang2023}~\cite{Fang2023} by applying a uniform singular value amplification~\cite{Gilyen2019} at each time step to bound the success probability, though at the cost of a quadratic dependence on the simulation time.

Here, a quantum algorithm for solving the advection-diffusion and heat equations using a direct, $\alpha=1$ block encoding is presented. A quantum algorithm for solving the advection equation \cite{Brearley2024} is combined with corrective unitary operators in the LCU framework \cite{Childs2012} to recover the advection-diffusion time-marching operator. The intrinsic success probability of the squared norm of the solution vector is achieved while retaining a linear simulation time dependence. This is beneficial since the success probability is solely dependent on the physical problem rather than the number of time steps. This improves over the general strategy in Eq.~\eqref{eq:general} \cite{Bharadwaj2024} in terms of the probability of success, and quantum singular value transformation methods \cite{Fang2023, Gilyen2019} in terms of the dependence on the simulation time.

\section{Algorithm}
The algorithm applies to multidimensional problems with spatially varying velocity fields. However, a one-dimensional periodic problem discretized using $N_\text{x}$ grid points with an equal grid spacing $\Delta x$ will briefly be considered for simplicity. The one-dimensional advection-diffusion equation discretized using the forward Euler method and a second-order central finite difference stencil in space is
\begin{equation}
\frac{\phi^m_{t+1}-\phi^m_t}{\Delta t} + v\frac{\phi^{m+1}_t - \phi^{m-1}_t}{2\Delta x} = D\frac{\phi^{m+1}_t - 2\phi^m_t + \phi^{m-1}_t}{(\Delta x)^2}\,,
\end{equation}
for discrete space and time locations $m$ and $t$, respectively. This produces the explicit time-marching equation
\begin{equation}
    \phi^{m}_{t+1} = \phi_t^{m-1}\left(r_\text{h} {+} \frac{r_\text{a}}{2}\right) + \phi_t^m\left( 1{-}2 r_\text{h} \right) + \phi_t^{m+1}\left(r_\text{h} {-}\frac{r_\text{a}}{2}\right)\,,
    \label{eq:time_marching_eq}
\end{equation}
where $r_\text{a} = v\Delta t/\Delta x$ and $r_\text{h} = D\Delta t/(\Delta x)^2$ are the stability parameters for advection and diffusion, respectively. Equation~\eqref{eq:time_marching_eq} can be written in the form of Eq.~\eqref{eq:explicit_time_stepping} with the time-marching operator
\begin{equation}
    A = 
    \begin{bmatrix}
        1{-}2 r_\text{h}    & r_\text{h}{-}\frac{r_\text{a}}{2}       &    &  0    & r_\text{h}{+}\frac{r_\text{a}}{2}   \\
        r_\text{h}{+}\frac{r_\text{a}}{2}  & \ddots    & \ddots    & &  0  \\
         & \ddots    & \ddots    & \ddots &  \\
        0 & & \ddots & \ddots & r_\text{h}{-}\frac{r_\text{a}}{2} \\
        r_\text{h}{-}\frac{r_\text{a}}{2}    &     0   &    & r_\text{h}{+}\frac{r_\text{a}}{2}        & 1{-}2r_\text{h}
    \end{bmatrix}\,.
    \label{eq:A_matrix}
\end{equation}%
Equation~\eqref{eq:explicit_time_stepping} is implemented efficiently by decomposing the matrix $A$ into an advection-like component and a corrective shift operator. The advection-like component can be implemented with the quantum advection algorithm of \citeauthor{Brearley2024}~\cite{Brearley2024}, which can then be linearly combined with the unitary shift operator using the LCU algorithm~\cite{Childs2012} to directly enact $A$ on a subspace of the quantum state. 

\subsection{Matrix decomposition}
\label{sec:matrix_decomposition}

The decomposition begins by temporarily scaling Eq.~\eqref{eq:A_matrix} to $A/(1-2r_\text{h})$ such that the diagonal entries are $1$, although this factor will later cancel out to recover the direct block encoding. Then, the scaled $A$ can be written as the sum
\begin{equation}
    \frac{A}{1-2r_\text{h}} = \hat{A} + \frac{2r_\text{h}}{1-2r_\text{h}}S \,,
    \label{eq:A_decomp}
\end{equation}
where
\begin{equation}
    \hat{A} =
    \begin{bmatrix}
        1    & \frac{-r_\text{h}-r_\text{a}/2}{1-2r_\text{h}}       &    &  0    & \frac{r_\text{h}+r_\text{a}/2}{1-2r_\text{h}}   \\
        \frac{r_\text{h}+r_\text{a}/2}{1-2r_\text{h}}  & \ddots    & \ddots    & &  0  \\
         & \ddots    & \ddots    & \ddots &  \\
        0 & & \ddots & \ddots & \frac{-r_\text{h}-r_\text{a}/2}{1-2r_\text{h}} \\
        \frac{-r_\text{h}-r_\text{a}/2}{1-2r_\text{h}}    &     0   &    & \frac{r_\text{h}+r_\text{a}/2}{1-2r_\text{h}}       & 1
    \end{bmatrix}\,
    \label{eq:A_adv}
\end{equation}
is the advection-like operator with the same mathematical structure as the time-marching operator for the discretized advection equation \cite{Brearley2024}, not to be confused with the advection-diffusion time marching operator $A$ from Eq.~\eqref{eq:A_matrix}, and
\begin{equation}
    S = \ket{N_\text{x}-1}\bra{0} + \sum_{j=1}^{N_\text{x}-1} \ket{j-1}\bra{j}\,
\end{equation}
is the unitary shift operator that shifts the quantum amplitudes to the preceding basis state.

\subsection{Quantum advection algorithm}
\label{sec:advection}

The constructed advection-like matrix $\hat{A}$ is implemented through a block encoding by Hamiltonian simulation \cite{Brearley2024, Gingrich2004}, which prepares a unitary $\text{e}^{-iH\theta}$ for a Hamiltonian evolution time per time step $\theta$. The Hamiltonian is defined as
\begin{equation}
    H =
    \begin{bmatrix}
        0 & -i\hat{A}^\dagger \\
        i\hat{A} & 0
    \end{bmatrix}\,,
\end{equation}
where $(\dots)^\dagger$ denotes the conjugate transpose, resulting in the unitary operator
\begin{equation}
 \text{e}^{-iH\theta}= {\exp}\begin{bmatrix}
    0 & -\hat{A}^\dagger \theta \\
    \hat{A} \theta & 0 
 \end{bmatrix}   \,.
 \label{eq:hamiltonian_simulation}
\end{equation}
\citeauthor{Brearley2024}~\cite{Brearley2024} showed that Eq.~\eqref{eq:hamiltonian_simulation} has the exact block matrix structure 
\begin{equation}
    e^{-iH\theta} = 
    \begin{bmatrix} 
        \cos(\sqrt{\hat{A}^\dagger \hat{A}}\,\theta) & 
        -\hat{A}^\dagger \frac{\sin(\sqrt{\hat{A}\hat{A}^\dagger} \, \theta)}{\sqrt{\hat{A}\hat{A}^\dagger}}
        \\ 
        \hat{A}\frac{\sin(\sqrt{\hat{A}^\dagger \hat{A}}\, \theta)}{\sqrt{\hat{A}^\dagger \hat{A}}} & 
        \cos(\sqrt{\hat{A}\hat{A}^\dagger }\, \theta) 
    \end{bmatrix} \label{eq:U_definition} \, ,
\end{equation}
in terms of matrix trigonometric functions. The unitary operator in Eq.~\eqref{eq:U_definition} can prepare the state $\hat{A}\ket{\psi}$ when $\theta=\pi/2$ with error $\mathcal{O}( [r_\text{a}+r_\text{h}]^2)$ \cite{Brearley2024}, conditional on measuring an ancilla qubit initialized as $\ket{0}$ in the state $\ket{1}$. The error of the encoding arises from the equivalent advection stability parameter of $\hat{A}$ being $(2r_\text{h}+{r_\text{a}})/(1-2r_\text{h})$, which has the asymptotic behavior of $\mathcal{O}(r_\text{a} + r_\text{h})$ in the limit of small $r_\text{h}$, and the advection algorithm can prepare a block encoding with an error that is the square of the equivalent advection stability parameter \cite{Brearley2024}. The encoding is direct when $\theta=\pi/2$ and postselection succeeds with near-certainty \cite{Brearley2024}.

\subsection{Linear combination of unitaries}
\label{sec:lcu}

\begin{figure*}
    \centering
    \includegraphics[width=\textwidth]{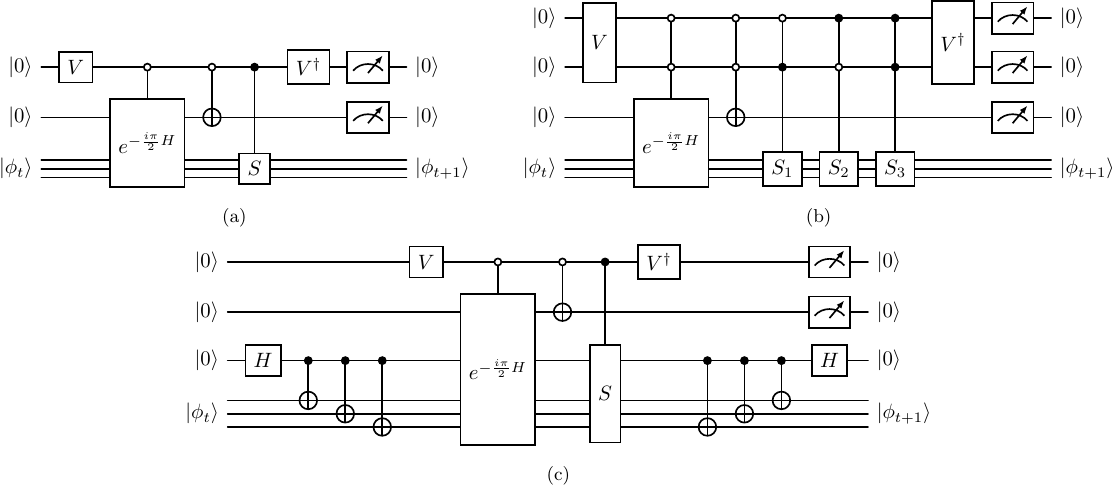}
    \caption{Quantum circuit for implementing a single time step of the algorithm in (a) one dimension ($d=1$), (b) three dimensions ($d=3$), and (c) one dimension ($d=1$) with homogeneous Neumann boundary conditions.}
    \label{fig:quantum_circuits}
\end{figure*}

A single time step for solving the one-dimensional advection-diffusion equation can be implemented using the quantum circuit shown in Fig.~\hyperref[fig:quantum_circuits]{\ref*{fig:quantum_circuits}a}, which uses the LCU algorithm \cite{Childs2012} to combine the quantum advection algorithm \cite{Brearley2024} with the shift operator. 

The unitary
\begin{equation}
    V = 
    \begin{bmatrix}
        \sqrt{\frac{\kappa_0}{\kappa_0+\kappa_1}} & -\sqrt{\frac{\kappa_1}{\kappa_0+\kappa_1}} \\
        \sqrt{\frac{\kappa_1}{\kappa_0+\kappa_1}} & \sqrt{\frac{\kappa_0}{\kappa_0+\kappa_1}}
    \end{bmatrix}
\end{equation}
encodes the square roots of the coefficients of the LCU sum in the first column corresponding to an ancilla measurement of $\ket{0}$, where $\kappa_0=1$ and $\kappa_1 = 2r_\text{h}/(1-2r_\text{h})$ from Eq.~\eqref{eq:A_decomp} for the one-dimensional problem briefly being considered. Implementing the circuit in Fig.~\hyperref[fig:quantum_circuits]{\ref*{fig:quantum_circuits}a} maps
\begin{equation}
    \begin{split}
            \ket{00}\ket{\phi_t} &\mapsto \ket{00}\frac{\kappa_0 \hat{A} + \kappa_1 S }{\kappa_0 + \kappa_1}\ket{\phi_t} + \dots \\
            &\mapsto \ket{00}A\ket{\phi_t} + \dots \, ,
            \label{eq:mapping_1d}
    \end{split}
\end{equation}
where $\hat{A}$ is represented with error $\mathcal{O}( [r_\text{a}+r_\text{h}]^2)$ from the full bottom-left element of the matrix in Eq.~\eqref{eq:U_definition}. The mapping simplifies to the action of the original, unscaled time-marching operator $A$ in Eq.~\eqref{eq:explicit_time_stepping}, since the normalization of $\kappa_0 + \kappa_1 = 1/(1-2r_\text{h})$ entirely cancels the original normalization of $1-2r_\text{h}$ in Eq.~\eqref{eq:A_decomp}. This property generalizes to higher dimensions as will be shown in Sec.~\ref{sec:generalization}, but is restricted to second-order central schemes in space. 

The probability of success for a time step is
\begin{equation}
    p_t = \bra{\phi_t}A^\dagger A\ket{\phi_t} = \frac{\| \vec{\phi}_{t+1} \|^2}{\| \vec{\phi}_{t} \|^2} \,,
    \label{eq:overall_probability} 
\end{equation}
where vector notation denotes the unnormalized solution. This results in a cumulative probability of $\|\vec{\phi}(T)\|^2 / \|\vec{\phi}(0) \|^2$ for the entire simulation time $T$. 
For a continuous variable $\phi(x,t)$ defined in $x\in\Omega$ by considering $\Delta x\to 0$, the probability of success becomes
\begin{equation}
    \frac{\int_{\Omega} |\phi(x,T)|^2 \, dx}{\int_{\Omega} |\phi(x,0)|^2 \, dx}\,,
    \label{eq:continuous_probability}
\end{equation}
which is intrinsic to the physical problem being solved rather than an artifact of numerical discretization.

The discussed probability of success can be improved to $\mathcal{O}(\|\vec{\phi}(T)\|/\|\vec{\phi}(0)\|)$ with uniform singular value amplification \cite{Gilyen2019} by removing the need for intermediate measurements \cite{Fang2023}. However, this is at the cost of a quadratic gate complexity in $T$, so its inclusion is therefore case-dependent.

\subsection{Generalization to higher dimensions}
\label{sec:generalization}

Applying the methodology to $d$~spatial dimensions requires a shift operator corresponding to each spatial dimension for a linear combination of $d+1$~unitary operators. The quantum circuit for $d=3$ is shown in Fig.~\hyperref[fig:quantum_circuits]{\ref*{fig:quantum_circuits}b}. The normalization in Eq.~\eqref{eq:A_decomp} generalizes to $1-2dr_\text{h}$. Because of the need to encode $d+1$ coefficients, the first column of the unitary $V$ now consists of
\begin{equation}
    V_{i,0} = \sqrt{\frac{\kappa_i}{\sum_{j=0}^d \kappa_j}}
\end{equation}
for rows $i=0$ to $d$, where $\kappa_0 = 1$ corresponds to the advection term, and $\kappa_1, \dots, \kappa_d = 2r_\text{h}/(1-2dr_\text{h})$ correspond to the $d$ spatial dimensions. Any excess coefficients, $\kappa_{d+1}$ onward, are set to zero with the corresponding shift operators excluded from the circuit. The remaining columns form an orthonormal basis such that $V$ is unitary. In $d$ dimensions, the quantum algorithm requires $n = \lceil \log_2(d+1)\rceil+1$ ancilla qubits and maps
\begin{equation}  
    \begin{split}
          \ket{0}^{\otimes n}\ket{\phi_t} &\mapsto \ket{0}^{\otimes n}\frac{\kappa_0 \hat{A} + \sum_{j=1}^d \kappa_jS_j}{\sum_{j=0}^d \kappa_j} \ket{\phi_t} + \dots \\
        &\mapsto \ket{0}^{\otimes n}A\ket{\phi_t} + \dots \, .
    \end{split}
\end{equation}
Evidently from the definitions of $\kappa_i$, the property of $\sum_{j=0}^d \kappa_j = 1/(1-2dr_\text{h})$ holds for all positive integers of $d$, ensuring that the intrinsic probability of success in Eqs.~\eqref{eq:overall_probability} and \eqref{eq:continuous_probability} applies in every multidimensional setting.

\subsection{Boundary conditions}
\label{sec:boundary}

The proposed algorithm can be extended to Neumann and Dirichlet boundary conditions by two distinct methodologies, each with their advantages and disadvantages. The first methodology is the simulation of an extended domain with even or odd symmetry at the boundaries, and the second methodology is the inclusion of additional unitaries into the LCU sum to modify the boundary values of $A$.

Neumann and Dirichlet boundary conditions can be simulated within a subspace of the periodic domain by reflection with even and odd symmetry at the subspace boundary, respectively. This method is widely used in the numerical simulation of PDEs using spectral methods, for example by the discrete sine transformation that naturally corresponds to a Dirichlet condition, or the discrete cosine transformation that naturally corresponds to a Neumann condition \cite{Canuto2007}. The operation is inherently unitary and requires one additional ancilla qubit per non-periodic spatial dimension to perform the reflection, as shown by \citeauthor{Sano2024}~\cite{Sano2024}. At the end of the simulation, the symmetric subdomains can simply be discarded by uncomputation. For example, the quantum circuit for implementing homogeneous Neumann boundary conditions for the one-dimensional heat equation is shown in Fig.~\hyperref[fig:quantum_circuits]{\ref*{fig:quantum_circuits}c}. The linear complexity in the simulation time is retained by this approach.

Alternatively, the methodology can be extended to non-periodic boundary conditions by including additional pairs of unitary operators into the LCU sum that, with the appropriate coefficients, adjust the boundary rows of $A$ to implement the desired condition. For example, implementing an insulated boundary corresponds to a homogeneous Neumann condition, and results in a first boundary row of $A$ of $[1{-}r_\text{h}, r_\text{h}, 0, \dots, 0]$ for the diffusion terms in one-dimension. This can be implemented using two pairs of unitary operators. The first pair contains an identity operator and a negative identity operator with boundary values of 1, which sum to a zero matrix with boundary values of 2. Therefore, including the weighting of $r_\text{h}/2$ in the LCU sum will add $r_\text{h}$ to the boundary value, effectively converting it from $1-2r_\text{h}$ to $1-{r_\text{h}}$. This must be repeated for the top-right and bottom-left corner elements of the matrix, by subtracting the $r_\text{h}$ term that implements periodicity. The inclusion of such unitaries for Dirichlet and Neumann conditions using this methodology does not retain the direct encoding described in the previous subsection. This can be overcome with the methodology of \citeauthor{Fang2023}~\cite{Fang2023} by applying a {uniform singular value amplification}~\cite{Gilyen2019} to remove the unwanted $\alpha>1$ factor. Hence, the above methodology with non-periodic boundary conditions can be considered as a specific implementation of the algorithm by \citeauthor{Fang2023} \cite{Fang2023}. This approach's trade-off for non-periodic problems is a quadratic scaling in simulation time. Since the method of reflection discussed in the previous paragraph retains the linear complexity in the simulation time, it will be superior for most applications.

\section{Complexity}

Consider the simulation of a $d$-dimensional spatial domain discretized with $N=N_\text{x}^d$ grid points. The simulation is evolved over a time interval $T=N_\text{T}\Delta t$ using a number of time steps $N_\text{T}$ with a constant time step $\Delta t$. This requires $\mathcal{O}(\log N + \log d)$ qubits, $\widetilde{\mathcal{O}}(TN^{2/d}d/\epsilon)$ two-qubit gates, and $\mathcal{O}(\|\vec{\phi}(0)\|^2 / \|\vec{\phi}(T)\|^2)$ attempts for a successful run, where $\widetilde{\mathcal{O}}$ is the complexity with the suppression of polylogarithmic terms and $\vec{\phi}(t)$ is the unnormalized solution at time~$t$. The probability of success and qubits requirements were demonstrated in Sec.~\ref{sec:lcu} and Sec.~\ref{sec:generalization} respectively, so now the two-qubit gate complexity will be derived. The algorithm requires the repeated application of the quantum circuits in Fig.~\ref{fig:quantum_circuits} for the desired number of time steps, so the gate complexity grows linearly with the desired number of time steps $N_T = \mathcal{O}(TN^{2/d}/\epsilon)$. The number of time steps growing linearly with the desired simulation time $T$ is apparent.
The $N^{2/d}$ component arises from the limiting stability condition $r_\text{h}$ that constrains $\Delta t \propto (\Delta x)^2$ and $N_\text{T} \propto N_\text{x}^2$, and since $N=N_\text{x}^d$, then $N_\text{T}\propto N^{2/d}$.
The $1/\epsilon$ component arises from the inverse relationship $N_\text{T}\propto 1/\Delta t$ and that $\epsilon \propto \Delta t$, resulting from the error $\epsilon$ of the underlying Euler method and the accuracy to which $\hat{A}$ is block-encoded using the advection algorithm \cite{Brearley2024}.
Each time step requires the implementation of a Hamiltonian simulation, where optimal algorithms \cite{Low2019} require a number of two-qubit gates that grow linearly with the sparsity of the Hamiltonian $H$, the largest absolute element of $H$ and the evolution time. The algorithms are at least polylogarithmic in other parameters, and these terms will be omitted through the $\widetilde{\mathcal{O}}$ notation.
In the context of the Hamiltonian simulation in Sec.~\ref{sec:advection}, the sparsity is $s=1+2d=\mathcal{O}(d)$ from the finite difference method, the largest absolute element $\|H\|_\text{max}=1$ and the evolution time $\theta=\pi/2$, which are both $\mathcal{O}(1)$. This completes the derivation of the overall two-qubit gate complexity of $\widetilde{\mathcal{O}}(TN^{2/d}d/\epsilon)$.
This contrasts with the typical classical complexity of $\mathcal{O}(NN_\text{T}) = \mathcal{O}(TN^{(2+d)/d})$, where the classical error dependence has been omitted due to the prevalence of high-order numerical methods in classical computing \cite{Laizet2009}. Comparing the expressions, the quantum algorithm provides a polynomial speed-up by reducing the exponent of $N$ by a factor of $2/(2+d)$.

\section{Simulations}
\label{sec:simulations}

\begin{figure}
    \centering
    \includegraphics[width=\columnwidth]{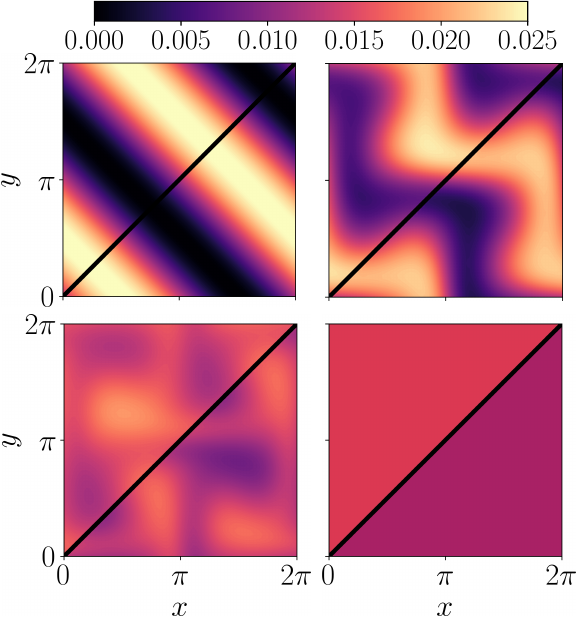}
    \caption{Contours of the scalar evolving by state-vector simulation (normalized, top-left half) and the corresponding classical solution (unnormalized, bottom-right half) for $t/T= 0$, $0.1$, $0.2$ and $1$ in left-to-right, top-to-bottom order.}
    \label{fig:contours}
\end{figure}

\begin{figure}
    \centering
    \includegraphics[width=\columnwidth]{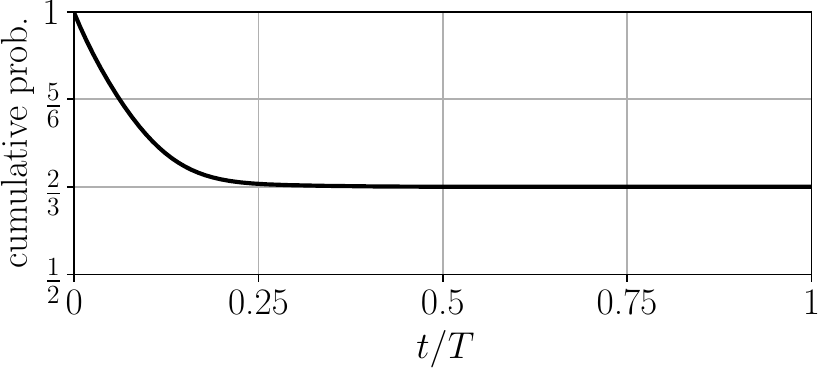}
    \caption{Cumulative probability of algorithm success, equivalent to $\|\vec{\phi}(t)\|^2/\|\vec{\phi}(0)\|^2$, demonstrating convergence to the theoretical value of $2/3$.}
    \label{fig:probability}
\end{figure}

In this section, the quantum algorithm is applied to state-vector simulations of passive scalar transport in a steady two-dimensional flow field that has characteristics of a Taylor-Green vortex with velocity
\begin{align}
\vec{v} (x,y) = 
\begin{bmatrix}
    \sin(x)\cos(y) \\
    -\cos(x)\sin(y)
\end{bmatrix} 
\end{align}
in a periodic domain $x,y\in [0, 2\pi)$. The domain is discretized by $64\times 64 = 4096$ grid points corresponding to 15 qubits including three ancilla qubits. The simulation time $T$ is divided into $N_\text{T} = 1400$ time steps with the maximum advection stability parameter $\max(r_\text{a})=\max(|v_x| + |v_y|)\Delta t/\Delta x =0.1$ and the uniform diffusion stability parameter $r_\text{h}=0.1$. This results in an adequate temporal resolution by ensuring at least 10 time steps are taken for the fluid to traverse the spatial resolution $\Delta x$ by advection, and a diffusion stability parameter that satisfies the condition of $r_\text{h}<1/(2d)$, derived from Fourier-von Neumann stability analysis \cite{Charney1950}. It was shown by \citeauthor{Brearley2024}~\cite{Brearley2024} that $r_\text{a}$ is not subject to the conditional stability of explicit methods ($r_\text{a} \leq 1$) due to the bounding of the spectral radius in the block encoding of a non-unitary operator, although the condition is met regardless. The scalar $\phi(x,y)$ is initialized proportionally to $\sin(x+y)+1$ with a norm of 1, defined as a non-negative function of both spatial coordinates to be an effective tracer of the two-dimensional velocity field, which may represent a physical quantity such as a species concentration. Figure~\ref{fig:contours} shows the evolution of $\phi(x,y)$ by the described quantum algorithm and by the classical action of $A$ directly. The classical simulation converges to a constant steady state of $\text{mean}(\vec{\phi}(0))$, while the quantum simulation converges to $1/\sqrt{N}$ due to the requirements of norm preservation. The P\'{e}clet number, defined as the ratio of the advective to diffusive transport, is $\text{Pe}= \vec{v}_{\text{rms}} \pi/D=23$ for root-mean-square velocity $\vec{v}_\text{rms} = 1/\sqrt{2}$ and characteristic length $\pi$, indicating advection as the primary mode of scalar transport with the influence of diffusion occurring over longer time scales. The quantum algorithm closely approximates the action of $A$ with a mean-squared error compared against the normalized classical solution not exceeding $0.5\%$, given as a percentage of $\max(|\vec{\phi}_t|^2)$. This is comparable to other more general algorithms for solving the advection-diffusion equation \cite{Ingelmann2024}. Figure~\ref{fig:probability} shows the cumulative probability of measurement success for the simulation, demonstrating that the algorithm success probability approaches the intrinsic value of $\| \vec{\phi}(T) \|^2 / \|\vec{\phi}(0) \|^2 = 2/3$ for these initial conditions.

\section{Conclusion}
The presented quantum algorithm simplifies the solution of diffusion and scalar transport problems by providing an accurate, direct block encoding of the explicit time-marching operator, achieving the intrinsic success probability of the squared solution norm. 

For periodic computational domains, the algorithm achieves a linear simulation time dependence, improving over other algorithms such as the linear combination of Hamiltonian simulation \cite{An2023} and time-marching algorithms using amplitude amplification \cite{Fang2023}, while retaining their advantageous non-vanishing success probabilities. The proposed algorithm offers an improvement over approaches based on quantum linear systems solvers, e.g., by \citeauthor{Berry2017}~\cite{Berry2017}, as there is no $\epsilon$ dependence in the number of queries to the state initialization oracle~\cite{Costa2022}. Simulating $d$-dimensional problems requires a linear combination of $d+1$ unitary operators to recover the unscaled time-marching operator, improving over general methods requiring four unitaries that asymptotically approach the target matrix \cite{Bharadwaj2024}, with diminishing success probabilities over successive time steps. Moreover, the presented quantum algorithm is efficient in its usage of ancillary qubits, with requirements that increase logarithmically with the number of spatial dimensions and independently of the simulation time, unlike the other approaches \cite{An2023,Fang2023}.

Non-periodic boundaries can be implemented by including additional unitary operators into the LCU circuit (see Sec.~\ref{sec:boundary}). The algorithm, combined with amplitude amplification \cite{Gilyen2019}, then becomes a specific implementation of the general time-marching algorithm \cite{Fang2023} for advection-diffusion problems. The probability of success improves to $\mathcal{O}(\|\vec{\phi}(T)\| / \|\vec{\phi}(0)\|)$, but at the cost of a quadratic $\mathcal{O}(T^2)$ scaling in the evolution time. Alternatively, non-periodic conditions can be implemented by introducing symmetry along the computational boundary, with parity that is dependent on the specific boundary condition. This approach retains the linear $\mathcal{O}(T)$ scaling in the evolution time. 

The algorithm is limited to the forward Euler method with a second-order central discretization of the spatial derivatives, although the latter is not so restrictive given the capacity to simulate extremely large computational domains on quantum computers. 

The general methodology of matrix decomposition, as presented in Sec.~\ref{sec:matrix_decomposition}, may have varied applications extending beyond scalar transport problems. As a potential candidate due to its similar mathematical structure, the Fokker-Planck equation opens up potential applications in stochastic fields such as financial modeling, statistical mechanics, and chemical kinetics. Exploring these diverse applications represents an interesting premise for future research.

\begin{acknowledgments}
    PO, SB, and TR are supported by the European Union's Horizon Europe research and innovation program (HORIZON-CL4-2021-DIGITAL-EMERGING-02-10) under grant agreement No.~101080085 QCFD. PB and SL are supported by EPSRC grant EP/W032643/1.
\end{acknowledgments}

\bibliography{apssamp}

\end{document}